# The interaction between trade and FDI: the CEE countries experience

Claudiu Tiberiu ALBULESCU[a,b,♣] and Daniel GOYEAU[b]

[a] Management Department, Politehnica University of Timisoara, P-ta. Victoriei, No. 2, 300006, Timisoara, Romania

[b] CRIEF, University of Poitiers, 2, rue Jean Carbonnier, 86022 Poitiers Cedex, France

**Abstract:**

Inside the EU, the commercial integration of the CEE countries has gained remarkable momentum before the crisis appearance, but it has slightly slowed down afterwards. Consequently, the interest in identifying the factors supporting the commercial integration process is high. Recent findings in the new trade theory suggest that FDI influence the trade intensity but the studies approaching this relationship for the CEE countries present mixed evidence, and investigate the commercial integration of CEE countries with the old EU members. Against this background, the purpose of this paper is to assess the CEE countries' intra-integration, focusing on the Czech Republic, Hungary, Poland and the Slovak Republic. For each country we employ a panel gravitational model for the bilateral trade and FDI, considering its interactions with the other three countries in the sample on the one hand, and with the three EU main commercial partners on the other hand. We investigate different facets of the trade – FDI nexus, resorting to a fixed effects model, a random effects model, as well as to an instrumental variable estimator, over the period 2000-2013. Our results suggest that outward FDI sustains the CEE countries' commercial integration, while inward FDI has no significant effect. In all the cases a complementarity effect between trade and FDI is documented, which is stronger for the CEE countries' historical trade partners. Consequently, these findings show that CEE countries' policymakers are interested in encouraging the outward FDI toward their neighbour countries in order to increase the commercial integration.

---

[♣] Corresponding author. E-mail: claudiu.albulescu@upt.ro. Tel: 0040-743-089-759; Fax: 0040-256-404-287

# 1. Introduction

Before the crisis appearance, the economy of the Central and Eastern European (CEE) countries has been immersed in an accelerated process of commercial integration. At the same time, the removal of barriers to capital flows has encouraged the regional outward and inward foreign direct investment (FDI). However, the crisis has slightly slowed down the CEE countries' real converge process, especially the bilateral investment. In this context, several simple questions arise: What is the influence of the trade – FDI interaction on the CEE countries' commercial integration? Do we notice a stabilisation of the CEE countries' commercial integration at long-run levels? Does the commercial integration with the new and old EU members present similar features for the CEE countries?

The relationship between trade and FDI is analysed from the point of view of the substitution or the complementarity effect which may appear between them. Historically, based on the theory of multinational firms, the horizontal FDI is considered as an alternative way for firms to internationalise. Therefore, the substitution effect between trade and FDI prevails over complementarity when countries are similar in size, technologies and factor endowments (Markusen, 1984; Markusen and Venables, 1999; Türkcan, 2007). When focusing on vertical FDI linkages, the literature documents in general a complementarity effect between trade and FDI (Helpman, 1984; Clausing, 2000). Recent developments in the new trade theory also emphasise that trade and FDI can be positively correlated (Lipsey and Weiss, 1984; Fontagné, 1999; Fontagné and Pajot, 1999).

At the European Union (EU) level, most of the papers focus on the commercial integration between the old EU members (Chiappini, 2011; Chiappini, 2013), or between the new EU members (Herrmann and Jochem, 2005; Kutan and Vukšić, 2007), and provide different results, depending on the empirical methodology and data employed. None of these studies investigates the particular case of the CEE countries' bilateral commercial integration. Thus, our paper brings several contributions to the existing literature.

First, our focus falls on testing the relationship between the bilateral trade and FDI considering the CEE countries, which recorded an increased regional commercial integration before the crisis outburst. As far as we know, this is the first paper addressing the linkages between the bilateral trade and FDI in the CEE countries. Data availability for this group of countries, especially in the case of the bilateral FDI, makes the task difficult. We need to deal with a trade-off between focusing on specific countries for which data are in general available for a longer period, or including in the sample more CEE countries, but reducing thus the number of observations. We therefore chose the first option and we test the trade – FDI nexus for the Czech Republic, Hungary, Poland and the Slovak Republic (CEE-4), using Organisation for Economic Co-operation and Development (OECD) statistics for the period 2000-2013. The OECD statistics contain information about the bilateral investments between the OECD members. Six CEE countries are OECD members, namely the Czech Republic, Estonia, Hungary, Poland, the Slovak Republic and Slovenia. However, Slovenia and Estonia are excluded from our sample because of insufficient number of observations. Moreover, we are forced to start with the year 2000 as no previous data are available for the Slovak Republic's inward FDI before this year[1].

The selected countries (to a smaller extent Poland), represent the countries with the highest level of trade openness in the CEE group of countries. According to the World Bank statistics (World Development Indicators), over the period 2000-2013, the trade openness in CEE-4 is above the average level of the entire group of CEE countries (118% compared to 103%), and considerably above the EU level (72%). At the same time, CEE-4 is historically considered as representing the advanced group of CEE countries (together with Slovenia), where the commercial and capital flows are more developed as compared to other CEE countries.

The second contribution of the paper consists in drawing a comparison between the role of the bilateral FDI in explaining the trade for CEE-4 having as partners the CEE countries on the one hand (CEE-3)[2], and the three main EU commercial partners (Austria, Germany and Netherlands – EU-3)[3] on the other hand. Assuming a complementary effect between trade and FDI, authorities must pay special attention to the bilateral investment between the CEE countries only if it determines a stronger commercial integration as compared to the main EU partners. The same applies in the case of a substitution effect.

Third, we test the trade – FDI interactions in a panel gravitational framework, constructing a panel data for each CEE-4 country. Different from previous works which include in a single panel all the CEE countries and which consider a group of countries (usually the old EU members) as partners, we focus on each CEE country for two reasons. On the one hand, there are important discrepancies between the data reported by different countries[4]. So, in order to obtain reliable results, we refer to the statistics reported by each country. On the other hand, building a panel for each CEE-4, allows us to make comparisons between these countries. Consequently, for each panel (corresponding to each CEE-4 country), we have as partners the other three CEE countries (CEE-3) and the three main EU commercial partners (EU-3). In order to make a comparison between the effect of FDI on trade in the case of CEE-3 and EU-3, we use a dummy variable. The empirical approach consists in drawing a comparison between a panel fixed effects model and a random effects model (classical approaches for dealing with macro panels). We also treat the endogeneity problems which may appear between trade and FDI, relying on an instrumental variable (IV) approach.

---

[1] Indeed, alternative databases such as United Nations Conference on Trade and Development (UNCTAD) are richer in bilateral FDI statistics. Nevertheless, there is a large heterogeneity between the CEE countries regarding the data availability, and for a large number of countries data are provided only after 2007.
[2] CEE-3 denotes the other three CEE countries acting as partners. For example, the CEE-3 in the case of the Czech Republic, are Hungary, Poland and the Slovak Republic.
[3] The three main commercial partners of the CEE-4, based on the trade volume (UNCTAD data).

Finally, we contribute to the literature exploring different facets of the trade – FDI relationship. Most of the existing works pay attention to the role of inward FDI for the commercial integration. Following the Fontagné (1999) approach, we test for the role of both inward and outward FDI on trade, considering the volume of exports and imports. This way we are able to identify different transmission channels from FDI to trade, and to better understand the nature of this relation.

Our results show a complementary effect between trade and FDI in the case of CEE-4. We document an important role of the outward FDI in favouring trade, while the role of the inward FDI is inconclusive. In addition, we find that this relation is stronger for the CEE-3 partners as compared to the EU-3 partners.

The remainder of the paper is organised as follows. Section 2 presents a short review of the literature. Section 3 describes the CEECs' commercial features, the data and the methodology. Section 4 presents the empirical results. Section 5 brings forward the conclusions and the policy implications of our findings.

## 2. Literature review

During the last decades, the world economy, and especially the EU economy, has become more integrated. The relationship between trade and FDI is considered as being at the core of this integration process. Consequently, the impact of FDI on trade has been intensely debated in the literature at macro level, since it provides information about the international specialisation of countries and about the general welfare effects. At the same time, at micro level, the firms' decisions to expand their markets encompass a combination of both trade and foreign direct investment, simultaneously determined by factors such as economies of scale, trade costs, market access and differences in factors endowments.

---

[4] For example, there are noticeable differences between the bilateral inward FDI reported by one country and the bilateral outward FDI reported by the partner country. At the same time, the exports reported by one country differ from the imports reported by the partner country. These discrepancies can be explained by different methodologies used. The exports are registered as FOB (*free on board*), while the imports as CIF (*cost, insurance, freight*). Another explanation regarding the statistical discrepancies is related to the exchange rate differential between the domestic currencies and the international currency used for transactions.

The trade – FDI nexus is examined both by the theories of international trade and by those of multinational companies. These theories, with an independent evolution, have emerged during the last years. Therefore, the common question nowadays is whether trade and FDI act as complements or substitutes in delivering goods across borders, whereas the answers to this question are quite divers. Overall, the new trade theory reveals that, depending on the circumstances, the FDI can have both a substitution[5], as well as a complementarity[6] effect on trade. For example, relying on the trade theory, Markusen (1997) and Carr et al. (2001) admit the complementarity, as well as the substitution, between FDI and trade. Based on the theory of firms' location, Pontes (2004) and Africano and Magalhães (2005) show that the complementarity between trade and FDI is normally found when foreign investments are vertical, meaning that the multinational companies split the production process across countries in order to reduce costs. At the same time, FDI substitutes trade when investments are horizontal.

However, the influence of FDI on trade is complex and it can be analysed from the perspective of the home or of the host country, but also from the point of view of the inward and outward FDI[7]. From the point of view of the home country, FDI is seen as substitute for trade, as exports are replaced by local sales on foreign markets, particularly in the form of final goods. For the host countries, the relationship between FDI and trade can be considered symmetrical to that of the investing country. The difference in factor endowment also explains the linkages between trade and FDI (see Helpman, 1984). At the same time, we can distinguish between the influence of the inward and outward FDI on trade, which needs not to be symmetrical. Therefore, Fontagné (1999) underlines four situations:

- Inward FDI influences exports if foreign firms locate in the host economy to export back home, or provide products/services in a regional market.

- Outward FDI influences exports owing to enhanced competitiveness on foreign markets or reduces exports if the opposite applies.

- Inward FDI influences imports owing to enhanced competitiveness of foreign firms on the domestic market, but they may give rise to exports when the host country gains competitiveness.

- Outward FDI influences imports in the case of backward vertical integration and/or relocation of labour-intensive activities abroad, from a capital-intensive country.

---

[5] The substitution effect is traditionally advanced by the international trade theory. As Pain and Wakelin (1998) show, in the conventional trade models based on the Heckscher-Ohlin-Samuelson (H-O-S) framework, the equalisation of factor prices across countries can be achieved either through the international trade, or through the international mobility of factors. In the latter case, factor mobility may be a substitute for trade if the production functions are identical (Mundell, 1957).
[6] The theory of multinational companies shows that through direct investments, these companies exploit natural resources which are not available in the home country. These investments are then more likely to create trade, by raising exports of capital equipment and factor services from the home country and exports of resource-based products from the host economy. Therefore, the trade and FDI are considered alternative means for entering foreign markets, underlying their complementarity (Caves, 1982).
[7] The trade – FDI nexus is also influenced by the effect of FDI on the economic activity and welfare in general. There are numerous benefits and costs associated with the FDI entrance (Markusen and Venables, 1999). While the benefits are associated with technological externalities, knowledge spillover and demonstration effects which may foster the trade, the costs are associated with the interaction between multinational companies and fixed distortions in the economy, and with additional competition. If for example the import tariff exceeds its optimal level because of the "tariff-hopping" FDI, the quantity of imports decreases.

The diversity of empirical works on the topic is really important. Most of the papers focus on FDI stocks (Kutan and Vukšić, 2007; Fillat-Castejón et al., 2008; Fontagné and Pajot, 1999; Chiappini, 2013), but there are also empirical works which take into account FDI flows (Zarotiadis and Mylonidis, 2005; Aydin, 2010). The use of FDI stocks is in general preferred due to smaller variations in data (FDI flows are extremely volatiles, especially in crisis periods) and especially due to the fact that the influence of FDI on trade is not instantaneous. Therefore, the cumulative role of the investments' flow, namely the stocks, must be considered (for more arguments see Bayoumi and Lipworth, 1997). If the majority of authors analyse the case of a single country (Pfaffermayr, 1994; Clausing, 2000; Alguacil and Orts, 2002), several recent papers use a panel data analysis (Kutan and Vukšić, 2007; Chiappini, 2011, 2013).

The fact that the theoretical and empirical literature exploits different issues and provides such different results does not stand for a surprise (Zarotiadis and Mylonidis, 2005). For example, in the case of the United States and Japan, Bayoumi and Lipworth (1997) find a substitutability relationship between trade and FDI, while Brainard (1997), Clausing (2000) and Co (1997) argue in favour of the complementarity between trade and FDI.

In the EU case, the link between trade and FDI is studied both from the single country perspective (Barry and Bradley, 1997; Africano and Magalhães, 2005; Zarotiadis and Mylonidis, 2005) and from that of a group of countries (Egger, 2001; Chiappini, 2011, 2013). If for example Zarotiadis and Mylonidis (2005) find for the United Kingdom, a complementarity effect of FDI on trade, Chiappini (2013) documents a strong complementarity relationship in the case of Germany, but not for France and Italy. Martínez et al. (2012) show in their turn that the EU commercial integration and the FDI level reinforce each other. Conversely, Egger (2001) documents a substitution effect between exports and stocks of outward FDI in the EU member states.

However, only a small number of papers investigate the case of the CEE countries, and only in relation with the old EU members. For example, Herrmann and Jochem (2005) use aggregate industry data, and find that FDI and trade are complements. In the same line, Sapienza (2009) tests the relationship between the bilateral flows of FDI and exports from the EU-15 towards the CEE countries, using an extended gravity approach that includes labour costs, and discovers a complementarity situation. Kutan and Vukšić (2007) analyse in their turn the potential effects of FDI inflows on exports in the case of twelve CEE countries, making the distinction between supply capacity-increasing effects and FDI-specific effects.

Nevertheless, none of the previous works analyses the effect of FDI on the CEE countries' intra-integration. In order to contribute to the empirical literature on this subject, we test the relationship between FDI and the bilateral trade in the case of four OECD countries, namely the Czech Republic, Hungary, Poland and the Slovak Republic, for the period 2000-2013.

## 3. Trade and FDI in the CEECs

### 3.1. Stylised facts

Until 1989, the CEE countries were planned economies with a trade organization based on the state's monopoly. The trade was characterised by a strong concentration inside the Council for Mutual Economic Assistance (CMEA). After 1989, during the structural reform period, the CEE countries' trade pattern has experienced significant changes (Guerrieri, 1998; Rault at al., 2007). We have noticed a strong expansion of the trade, in particular with the EU countries, whereas the CMEA intra-regional trade collapsed.

Three countries are to be noticed in that period, namely the Czech Republic, Hungary and Poland. They represent almost two-thirds of the trade volume with the old EU members (EU-15) in the 1990s. The Slovak Republic can also be added to this group (CEE-4) due to its strong historical relationship with the Czech Republic and to the high level of trade openness. The new policy configuration favoured the commercial integration between the CEE-4 and the EU-15. The commercial relationships with the old EU members became more and more important, with a growing trend situated well above the commercial trade between the CEE countries and the rest of the world, or between the CEE countries and the traditional Eastern European partners.

However, even if the CMEA disappeared, we remark a revival of the commercial trade between the CEE countries during the years preceding the crisis. Nevertheless, Fig 1 shows that, with the onset of the crisis, the commercial integration between the CEE-4 regressed in 2009, recovered immediately after and slowed down during the last years.

**Fig 1 Bilateral trade between each CEE-4 country and the CEE-3 partners (stocks, mil. US dollars)**

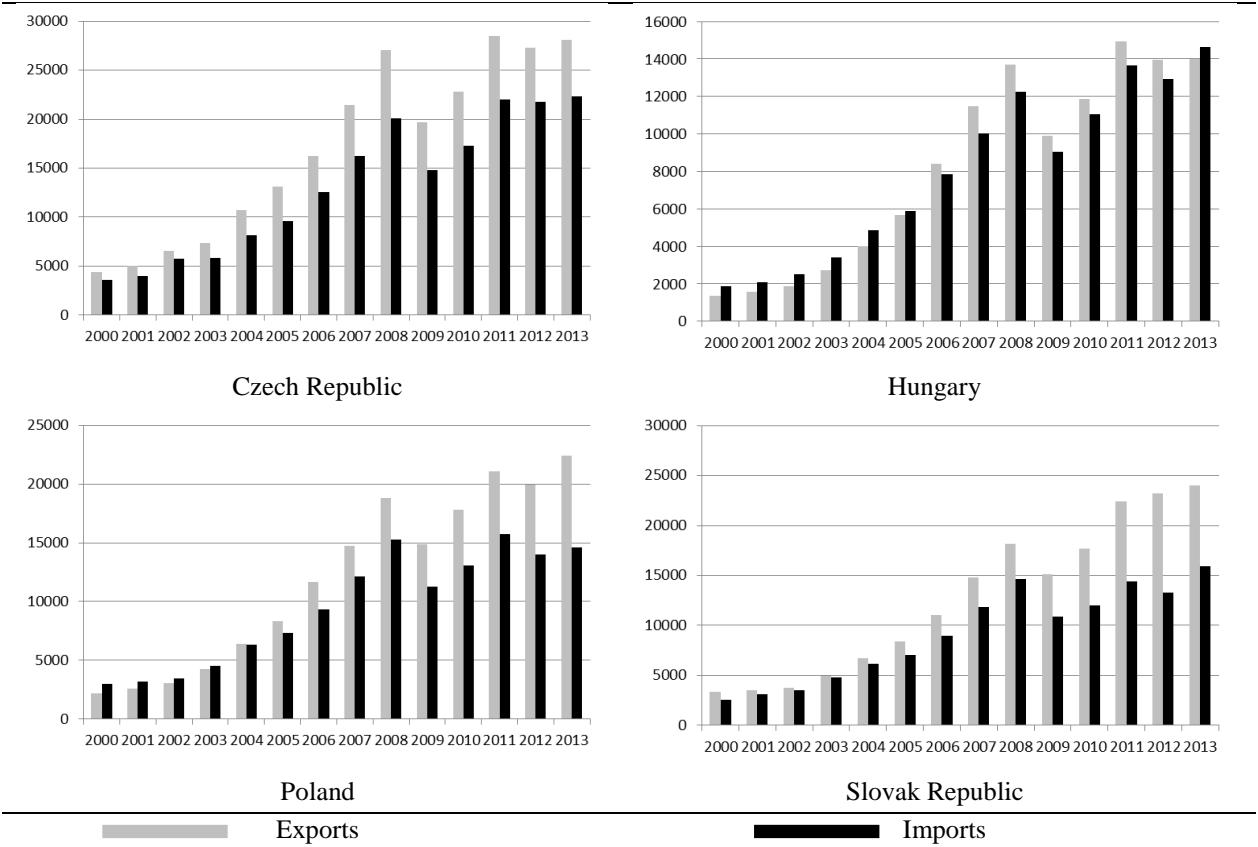

*Source: own calculations based on the OECD statistics*

According to the OECD data, in 2000 the bilateral trade in the case of the four selected countries represented less than 10% out of their total trade (world as partner). Ten years later, this percentage doubled. During the financial crisis period, the bilateral trade of the CEE-4 registered a slight contraction as compared to the overall trade activity of these countries (except for Hungary, where the bilateral trade with the CEE-3 countries increased as compared to its total trade volume).

The analysis of FDI reveals similar features (Fig 2 – Appendix A). Both inward and outward FDI (CEE-3 as partners) sharply increased during the last decade. The FDI volume multiplied five times in the analysed period, indicating a link with the bilateral trade intensification. If we compare the situation of each country, we notice that the Czech and the Slovak Republic are the destination countries of FDI from the CEE-3 partners. An opposite situation appears in the case of Poland and Hungary, where the inward FDI from the other three partners retained into the analysis is much more reduced as compared to the outward FDI toward these countries. Further, in all the countries, the bilateral outward FDI decreased after 2005, as compared to the total FDI stock (world as partner). We also notice that the outward FDI toward CEE-3, as compared to the total outward FDI in CEE countries is several times more important that the inward FDI, proving that CEE countries are FDI destination countries, while their investment abroad are mainly performed in the neighbouring countries.

### *3.2. Data description and methodology*

The annual data for the bilateral trade are extracted form UNCATD database while the bilateral FDI statistics are extracted from the OECD database, covering the period 2000-2013[8] (data are expressed in US dollars and natural logarithm). Having in mind the historical background of the commercial relationships between the CEECs-4, we are interested in exploring several aspects of the trade – FDI nexus. First, we want to see if FDI enhances the competitiveness of the host country. Consequently, we test the relation between outward FDI and exports. Second, we want to see if FDI is accompanied by an increasing volume of factors from the investor countries, factors supporting the production process. These factors are meant to increase the competitiveness of foreign firms on the domestic market. Therefore, we test the relation between inward FDI and imports. Third, we analyse the backward vertical integration and the relocation abroad of labour-intensive activities. We thus analyse the impact of outward FDI on imports. Finally, we look for "export-back effect" and we test the relationship between inward FDI and exports.

To validate our results we use a large battery of control variables (all variables are expressed in natural logarithm). The general equation we propose is largely employed in the bilateral trade models with factor endowments (Herrmann and Jochem, 2005; Fontagné and Pajot, 1999; Egger, 2002; Chiappini, 2013). The bilateral trade is influenced by the bilateral income, the country relative size, the difference in factor endowments and trade costs (associated with the geographical distance between countries). A description of the explanatory variables is presented in Table 1 (Appendix B)[9].

---

[8] For the Slovak Republic, the FDI data for 2013 were extracted from the Investment Climate Statements 2014 (Bureau of Budget and Planning, US Department of State), as they were not available in the OCDE statistics. For the Czech Republic, the FDI data for 2002 are not available (we use linear interpolation to avoid broken panel problems). In line with most of previous works we use stock data (see the related arguments in the Literature review section).

The empirical methodology refers to a macro panel data analysis and a gravity approach. Two reasons underlie the choice of the panel data approach. First, the trade and FDI levels have both risen over time, due to generally better economic conditions. It thus makes little sense to investigate the impact of FDI on trade over time in isolation (Pain and Wakelin, 1998)[10]. Second, the lack of historical statistical data and their low frequency in the case of the CEE countries recommends the use of a panel analysis. We also employ a gravity equation which is widely and successfully used to explain bilateral trade flows. The trade patterns enable the analysis of the bilateral trade flow by taking in a simple manner into consideration few selected statistical data as the GDP, the geographical distance and a group of variables designed to capture specific aspects of the commercial integration of the considered countries (Sapienza, 2009)[11].

The gravity equations are usually estimated in panel with fixed effects in order to eliminate the problem of unobserved variables and to overcome the lack of control for the heterogeneous trading relationships. We start then our empirical investigation resorting to a fixed effects model.

$$trade_{ijt} = \beta_0 + \beta_1 fdi_{ijt} + \beta_2 X_{it} + \alpha_i + \varepsilon_{it} \qquad (1)$$

where: $trade_{ijt}$ is the bilateral trade between country $i$ and country $j$ at date $t$ (exports and imports); $fdi_{ijt}$ is the bilateral FDI stock between country $i$ and country $j$ at date $t$; $X_{it}$ represents the set of control variables; $\beta_0$ is the intercept; $\alpha_i$ represents all the stable characteristics of countries; $\varepsilon_{it}$ are the errors of the model.

However, this method does not control for all stable covariates (Allison and Waterman, 2002). Therefore, we also test a random model and we compare the fixed and random effects models based on the Hausman test.

$$trade_{ijt} = \beta_0 + \beta_1 fdi_{ijt} + \beta_2 X_{it} + \alpha_i + \mu_{it} + \varepsilon_{it} \qquad (2)$$

where: $\mu_{ijt}$ represents between-entity errors; $\varepsilon_{it}$ are the within-entity errors of the model.

Fontagné (1999) shows that trade may also cause FDI. Therefore, the endogeneity issue which arises between trade and FDI is not addressed by the previous models.

---

[9] The correlation matrix, computed for each panel, can be provided by the authors, under request.
[10] As Rault et al. (2007) state, the standard cross-section estimates tend to ignore the unobservable characteristics of bilateral trade relationships (historical, cultural and linguistic links).
[11] In its simplest form, the gravity equation states that the volume of trade between any two countries is positively correlated with the economic size of these countries and negatively correlated with the geographic distance between them (Martínez et al., 2012).

Consequently, we continue our analysis with an instrumental variable regression. We use a Two Stages Least Square estimator (2SLS) and we perform two endogeneity tests, namely the Wu-Hausman and the Durbin-Wu-Hausman tests. If the endogeneity is present, then the Ordinary Least Square (OLS) estimator is inconsistent and an IV approach is recommended (Baum et al., 2003). We also check for the presence of heteroskedasticity using the Pagan-Hall general test. According to Baum et al. (2007), if the heteroskedasticity problem exists, it can be corrected using robust standard errors. Given the small number of observations, in the case of the 2SLS approach, usually the proliferation of instruments constitutes a problem. Therefore we use the Sargan over-identification test for all instruments (the Hansen J test is recommended if robust standard errors are used).

Before performing the estimations, we check for the stationarity of our panel data. The common tests employed in the literature as the MW test (Maddala and Wu, 1999), the Choi test (Choi, 2001), the LLC test (Levin et al., 2002) and the IPS test (Im et al., 2003), are based on the assumption of independent cross-section units. Moreover, their asymptotical properties require large data samples (N, T →∞). However, in the case of macro panels, the cross-sectional independence hypothesis is strong and must be checked. We start thus with a series of cross-sectional dependence tests in order to verify this hypothesis (Friedman, 1937; Frees, 1995; Pesaran, 2004). The results of the cross-sectional dependence tests for each panel (and each tested relationship between trade and FDI) are presented in Table 3.

**Table 3. Cross-sectional dependence tests**

| Tests | Czech Republic | Hungary | Poland | Slovak Republic |
|---|---|---|---|---|
| *Exports – outward FDI* | | | | |
| Pearson CD Normal (Pesaran, 2004) | 5.060 (0.000) | 1.065 (0.286) | 3.674 (0.000) | 2.769 (0.005) |
| Friedman Chi-square (Friedman, 1937) | 32.70 (0.000) | 17.25 (0.004) | 30.43 (0.000) | 23.88 (0.000) |
| Frees Normal (Frees, 1995) | 0.709 | -0.050 | 0.587 | 0.186 |
| *Exports – inward FDI* | | | | |
| Pearson CD Normal (Pesaran, 2004) | 4.278 (0.000) | 1.521 (0.128) | 6.283 (0.000) | 1.815 (0.069) |
| Friedman Chi-square (Friedman, 1937) | 32.11 (0.000) | 17.60 (0.003) | 42.13 (0.000) | 20.76 (0.000) |
| Frees Normal (Frees, 1995) | 0.564 | -0.033 | 1.154 | 0.442 |
| *Imports – outward FDI* | | | | |
| Pearson CD Normal (Pesaran, 2004) | 1.374 (0.169) | 1.935 (0.053) | 3.398 (0.000) | 5.486 (0.000) |
| Friedman Chi-square (Friedman, 1937) | 21.73 (0.000) | 17.08 (0.004) | 30.24 (0.000) | 32.81 (0.000) |
| Frees Normal (Frees, 1995) | 0.271 | 1.075 | 0.485 | 0.698 |
| *Imports – inward FDI* | | | | |
| Pearson CD Normal (Pesaran, 2004) | 1.972 (0.048) | 2.053 (0.040) | 5.775 (0.000) | 3.571 (0.000) |
| Friedman Chi-square (Friedman, 1937) | 22.93 (0.000) | 17.42 (0.003) | 39.48 (0.000) | 28.62 (0.000) |
| Frees Normal (Frees, 1995) | 0.526 | 0.950 | 0.972 | 0.405 |

*Note: (i) The null hypothesis for each tests is the cross-sectional independence; (ii) Test statistic are reported (p-values in brackets); (iii) The critical values for Frees' Q distribution are 0.184 (10%), 0.243 (5%) and 0.360 (1%).*

We notice that in almost all the cases the null of cross-sectional independence is rejected, which makes the use of the first generation of panel unit root tests inappropriate. However, in several cases, the hypothesis cannot be rejected: in the case of Hungary, the Frees Normal test and the Pearson CD Normal test admit the cross-sectional independence for the exports – FDI relationship, while in the case of the Slovak Republic the independence is accepted at 10% confidence level by the same tests.

We then use the Pesaran cross-sectional Augmented Dickey–Fuller (CADF) test, to check for the presence of panel unit roots. Pesaran (2007) advances a modified statistics based on the IPS test (Im et al., 2003), considering the average of the individual CADF. However, besides this new test from the second generation which does not assume the cross-sectional independence and can be performed for a limited number of observations, we also resort to the IPS test for comparison reasons and because the cross-sectional dependence tests (Table 3) do not reject the independence hypothesis in all the cases. The results of the panel unit root tests for all considered variables are presented in Table 4 below. The tests provide mixed evidence regarding the stationarity of our variables, underlining the differences between the results of the first and the second generation of panel unit root tests.

**Table 4. Panel unit root tests**

|  | Czech Republic | | Hungary | | Poland | | Slovak Republic | |
|---|---|---|---|---|---|---|---|---|
| **Variables** | CADF test | IPS test | CADF test | IPS test | CADF test | IPS test | CADF test | IPS test |
| *ex* | -1.382 | -2.131 | -1.958 | -1.614 | -2.134 | -1.699 | -3.223$^t$ | -1.419 |
|  | (0.77) | (0.05) | (0.28) | (0.30) | (0.17) | (0.21) | (0.01) | (0.50) |
| *im* | -2.842$^t$ | -1.860 | -1.920 | -1.557 | -1.553 | -1.371 | -3.092 | -1.882 |
|  | (0.09) | (0.12) | (0.31) | (0.34) | (0.63) | (0.52) | (0.00) | (0.11) |
| *outfdi* | -2.964 | -2.650 | -2.585 | -2.071$^t$ | -3.023 | -2.093 | -1.832 | -2.235 |
|  | (0.00) | (0.09) | (0.02) | (0.10) | (0.00) | (0.18) | (0.39) | (0.04) |
| *infdi* | -1.727 | -1.977 | -1.840 | -2.196 | -2.833 | -1.739 | -2.719 | -2.119 |
|  | (0.48) | (0.10) | (0.38) | (0.04) | (0.00) | (0.19) | (0.01) | (0.05) |
| *gdpav* | -2.241 | -1.524 | -2.063 | -1.745 | -1.412 | -3.214$^t$ | -2.484 | -1.343 |
|  | (0.11) | (0.39) | (0.21) | (0.21) | (0.78) | (0.00) | (0.04) | (0.60) |
| *gdpdif* | -0.988 | -0.864 | -1.088 | -1.922$^t$ | -0.389 | -2.353$^t$ | -1.866 | -0.981 |
|  | (0.94) | (0.95) | (0.91) | (0.16) | (0.99) | (0.01) | (0.36) | (0.91) |
| *gdpcav* | -1.626 | -1.759 | -1.341 | -1.890 | -2.395 | -1.183 | -2.093 | -1.427 |
|  | (0.57) | (0.18) | (0.49) | (0.07) | (0.06) | (0.75) | (0.19) | (0.48) |
| *gdpcdif* | -1.747 | -0.724 | -1.341 | -1.290 | -3.365$^t$ | -1.042 | -0.254 | -0.519 |
|  | (0.46) | (0.97) | (0.77) | (0.64) | (0.00) | (0.91) | (0.99) | (0.99) |
| *gdpg* | -2.083 | -3.018 | -1.901 | -2.943 | -1.498 | -3.036 | -1.686 | -2.905 |
|  | (0.20) | (0.00) | (0.33) | (0.00) | (0.68) | (0.00) | (0.52) | (0.00) |
| *popav* | -0.987 | -1.435$^t$ | -0.671 | -1.273 | -0.480 | -1.087 | -1.253 | -0.491 |
|  | (0.94) | (0.68) | (0.99) | (0.70) | (0.99) | (0.82) | (0.84) | (0.99) |
| *bexr* | -2.238 | -2.291$^t$ | -1.968 | -2.547$^t$ | 0.072 | -2.566$^t$ | -1.680 | -2.152$^t$ |
|  | (0.11) | (0.03) | (0.28) | (0.00) | (1.00) | (0.00) | (0.52) | (0.09) |

*Notes: (i) The null hypothesis for both tests is the presence of panel unit root; (ii) The cross-sectional ADF (CADF) test is proposed by Pesaran (2007) assuming cross-sectional dependence, while the IPS test is proposed by Im et al. (2003), assuming cross-sectional independence; (iii) t-bar is reported and the p-values are in brackets; (iv) 2 lags are used fort the CADF test; (v) $^t$ represents test performed with trend for achieving the stationarity (otherwise the constant is used).*

## 4. Empirical findings

The trade – FDI nexus is estimated using three different methods (Tables 5-8). Table 5 shows that the outward FDI has a positive effect on the level of exports, proving a complementarity effect between trade and FDI (similar findings are reported by Herrmann and Jochem, 2005). These results can be noticed for all CEE-4 countries but to a smaller extent for Poland, as the 2SLS approach indicates a non-significant, negative sign for the outward FDI. The results are robust regarding the empirical approaches used. In general, the Hausman test recommends the fixed effect model, except for Poland. The endogeneity tests show no endogeneity problems for the 2SLS estimation. However, the Pagan-Hall test underlines some concerns regarding heteroskedasticity issues for Poland and the Slovak Republic. Therefore, for these two countries we use robust standard errors in order to address these issues. The Hansen J test shows that the over-identification of instruments persists for these panels.

**Table 5. Exports and outward FDI relationship**

| exports | Czech Republic | | | Hungary | | | Poland | | | Slovak Republic | | |
|---|---|---|---|---|---|---|---|---|---|---|---|---|
| | FE | RE | 2SLS | FE | RE | 2SLS | FE | RE | 2SLS$^r$ | FE | RE | 2SLS$^r$ |
| c | -51.4 | -46.2*** | -48.0 | 35.5 | -83.2*** | - | 90.4 | -9.77 | -176 | -88.2** | -56.4*** | -40.4** |
| **outfdi** | 0.06*** | 0.10*** | 0.07* | 0.02 | 0.08*** | 0.17** | 0.10*** | 0.10*** | -0.36 | 0.05*** | 0.05*** | 0.13*** |
| gdpav | -1.83* | -0.25 | -0.80 | -4.26*** | -1.00 | 4.79*** | 3.43*** | 2.96*** | -0.27 | -2.18** | 0.70 | 1.66** |
| gdpdif | -0.20 | 0.17 | 0.03 | 0.30** | 0.33** | -0.20 | -0.02 | -0.01 | -0.17 | -0.58** | -1.16** | -1.00*** |
| gdpcav | 6.34*** | 4.95*** | 5.38** | 9.91*** | 7.09*** | 0.78* | 0.89 | 1.21* | 11.6* | 6.60*** | 4.41*** | 3.19*** |
| gdpcdif | 0.12 | 0.16 | 0.04 | -0.01 | 0.06** | 0.14** | 0.03 | 0.02 | -0.08 | 0.00 | -0.04 | -0.01 |
| gdpg | -0.03 | -0.02 | -0.43 | 0.05 | -0.04 | 0.81 | 0.09 | 0.10* | 0.05 | 0.07 | 0.08 | 0.06 |
| popav | 1.85 | 1.17*** | 1.49 | -4.36 | 2.54 | -3.07*** | -7.58 | -1.48** | 4.72 | 4.44* | 2.25*** | 1.10 |
| bexr | -0.94*** | 0.08 | 0.05 | 0.99*** | -0.08 | 0.44*** | -0.06 | 0.02 | 0.18 | -0.43** | -0.39*** | -0.24** |
| dist | - | -1.57*** | -1.56* | - | -1.12*** | -1.24*** | - | -0.34** | 0.13 | - | -0.50*** | -0.54*** |
| dummy | - | 2.00*** | 1.93 | - | 1.43*** | 2.83*** | - | 1.53*** | 5.39** | - | 0.52 | 0.95*** |
| Hausman test (recommended) | $P > \chi^2 = 0.00$ (Fixed) | | | $P > \chi^2 = 0.00$ (Fixed) | | | $P > \chi^2 = 0.97$ (Random) | | | $P > \chi^2 = 0.06$ (Fixed) | | |
| Pagan-Hall test | | 5.83 (0.82) | | | 1.10 (0.99) | | | 0.17 (1.00) | | | 0.17 (1.00) | |
| Wu-Hausman test | | 1.38 (0.25) | | | 1.67 (0.18) | | | -1.24 (1.00) | | | 1.13 (0.34) | |
| Durbin-Wu-Hausman test | | 4.72 (0.19) | | | 5.55 (0.13) | | | -4.76 (1.00) | | | 3.89 (0.27) | |
| Sargan test / Hansen J test | | 0.00 | | | 0.00 | | | 1.30 (0.51) | | | 2.12 (0.34) | |
| Observations | 84 | 84 | 83 | 84 | 84 | 83 | 84 | 84 | 83 | 84 | 84 | 83 |

*Notes: (i) \*, \*\*, \*\*\* means significance at 10 %, 5 % et 1 %; (ii) p-values are reported in brackets; (iii) The instruments used for the endogenous variables (outfdi, gdpg and bexr) are the first lags of these variables; (iv) Sargan statistic represents the over-identification test for all instruments and equals 0.00 if the equation is exactly identified. However, if the Pagan-Hall test documents the presence of heteroskedasticity, the Hansen J is recommended for identifying over-identification issues; (v) 2SLS$^r$ means 2SLS with robust errors if the heteroskedasticity is present; (vi) for endogeneity issues we resort to the Wu-Hausman and Durbin-Wu-Hausman tests; (vii) the dummy variable takes value1 for CEE-3 as partners, and 0 otherwise.*

If we look to the control variable, we notice that the GDP growth rate of the partner country does not explain the trade level (it is expected that the trade intensity grows in normal periods, while having an opposite trend in crisis times). The average GDP per capita positively influences the bilateral trade. At the same time, the distance between countries negatively affects their commercial integration. Furthermore, the coefficient of the dummy variable is positive and significant in general, showing that the link between trade and FDI is stronger in the case of the CEE countries as partners.

We further on test the "export-back effect" (Table 6).

**Table 6. Exports and inward FDI relationship**

| exports | Czech Republic | | | Hungary | | | Poland | | | Slovak Republic | | |
|---|---|---|---|---|---|---|---|---|---|---|---|---|
| | FE | RE | 2SLS | FE | RE | 2SLS$^r$ | FE | RE | 2SLS | FE | RE | 2SLS$^r$ |
| c | -127*** | -54.6*** | -56.8 | 61.0 | -77.0*** | -118*** | 79.7 | -38.2*** | -38.5 | -134*** | -81.2*** | -105*** |
| **infdi** | 0.01 | 0.04 | 0.00 | 0.01 | 0.06** | -0.03 | 0.08 | 0.08* | 0.13 | 0.13*** | 0.14*** | 0.40 |
| gdpav | -3.56*** | -1.09** | -1.49 | -4.28*** | -1.02 | -3.25 | 2.86*** | 2.16*** | 1.68 | -2.61*** | -2.01*** | -5.65 |
| gdpdif | -0.06 | 0.43** | 0.53 | 0.36*** | 0.59*** | 0.71*** | -0.07 | -0.06 | -0.02 | -0.80*** | -0.92*** | -0.32 |
| gdpcav | 7.98*** | 5.96*** | 6.44* | 9.95*** | 7.08*** | 9.99*** | 2.42** | 2.96*** | 2.73 | 6.36*** | 5.84*** | 6.81*** |
| gdpcdif | 0.07 | -0.05 | -0.12 | -0.02 | 0.05** | 0.03 | 0.02 | 0.01 | 0.08 | -0.00 | -0.03 | 0.02 |
| gdpg | -0.02 | -0.06 | -0.39 | 0.03 | -0.06 | -0.10 | 0.08 | 0.08 | -0.46 | 0.08 | 0.08 | 0.08 |
| popav | 6.75** | 1.59*** | 1.80 | -6.00* | 2.02 | 4.44* | -7.36 | -0.16 | 0.32 | 7.86*** | 4.75*** | 7.82** |
| bexr | -1.16*** | 0.08 | 0.04 | 0.95*** | -0.20 | -0.24 | -0.05 | 0.03 | 0.02 | -0.45** | -0.48*** | -0.49*** |
| dist | - | -1.42*** | -1.47* | - | -1.09*** | -1.06*** | - | -0.45** | -0.39 | - | -0.45*** | -0.42*** |
| dummy | - | 2.16*** | 2.04 | - | 1.51*** | 1.06* | - | 2.39*** | 2.51*** | - | 0.13 | -0.18 |
| Hausman test (recommended) | $P > \chi^2 = 0.00$ (Fixed) | | | $P > \chi^2 = 0.00$ (Fixed) | | | $P > \chi^2 = 0.97$ (Random) | | | $P > \chi^2 = 0.95$ (Random) | | |
| Pagan-Hall test | | | 8.78 (0.55) | | | 0.62 (1.00) | | | 1.61 (0.99) | | | 0.44 (1.00) |
| Wu-Hausman test | | | 0.65 (0.58) | | | 2.14 (0.10) | | | 0.98 (0.40) | | | 0.73 (0.53) |
| Durbin-Wu-Hausman test | | | 2.28 (0.51) | | | 6.97 (0.07) | | | 3.42 (0.33) | | | 2.58 (0.46) |
| Sargan test / Hansen J test | | | 0.00 | | | 8.52 (0.01) | | | 0.00 | | | 0.71 (0.70) |
| Observations | 84 | 84 | 83 | 84 | 84 | 83 | 84 | 84 | 83 | 84 | 84 | 83 |

*Notes: (i) \*, \*\*, \*\*\* means significance at 10 %, 5 % et 1 %; (ii) p-values are reported in brackets; (iii) The instruments used for the endogenous variables (infdi, gdpg and bexr) are the first lags of these variables; (iv) Sargan statistic represents the over-identification test for all instruments and equals 0.00 if the equation is exactly identified. However, if the Pagan-Hall test documents the presence of heteroskedasticity, the Hansen J is recommended for identifying over-identification issues; (v) 2SLS$^r$ means 2SLS with robust errors if the heteroskedasticity is present; (vi) for endogeneity issues we resort to the Wu-Hausman and Durbin-Wu-Hausman tests; (vii) the dummy variable takes value1 for CEE-3 as partners, and 0 otherwise.*

If the inward FDI favours exports, both a regional market and a complementarity effect appear. This phenomenon manifests in two cases: when the host country presents competitive advantages in terms of labour costs (it is not the case here), or when the host country disposes of necessary skills in some industries (more plausible in respect of the composition of our sample). However, according to our results, this effect manifests only in the case of the Slovak Republic (the fixed and random effects models), but it is not sustained by the 2SLS estimation. It seems that the CEE countries which receive investments from the considered partners do not necessarily export the final goods toward these countries.

We continue our analysis exploring the role of FDI in stimulating the level of imports. Table 7 presents the results for the influence of outward FDI. Three situations appear. First, in the case of the Czech Republic, a positive influence of outward FDI on imports is noticed, validated by all the tests, and showing its backward vertical integration, stronger for the CEE-3 as partners, as compared to EU-3. Second, the complementarity is also document for Hungary and Poland, but the results are not robust (while the 2SLS shows a significant and positive coefficient for Hungary, the results for Poland are sustained only by the fixed and random effects models). Third, no significant relationship is found for the Slovak Republic.

**Table 7. Imports and outward FDI relationship**

| *imports* | Czech Republic | | | Hungary | | | Poland | | | Slovak Republic | | |
|---|---|---|---|---|---|---|---|---|---|---|---|---|
| | FE | RE | 2SLS$^r$ | FE | RE | 2SLS$^r$ | FE | RE | 2SLS$^r$ | FE | RE | 2SLS$^r$ |
| c | -38.8 | -27.3*** | -19.9 | -582*** | -130*** | -123*** | -15.4 | 4.57 | -115* | 63.6 | -28.8*** | -40.7*** |
| **outfdi** | 0.05*** | 0.09*** | 0.13** | 0.00 | 0.02 | 0.09** | 0.08*** | 0.08*** | -0.26 | 0.02 | -0.00 | -0.09 |
| gdpav | -0.10 | 0.13 | 0.54 | -4.16** | -4.48** | -3.78*** | 2.07** | 3.05*** | 0.81 | -1.95 | 2.88*** | 2.10** |
| gdpdif | 0.08 | 0.41*** | 0.30* | -0.35* | 0.51*** | 0.25 | -0.07* | -0.06 | -0.18** | -0.92*** | -1.32*** | -1.38*** |
| gdpcav | 3.96*** | 3.96*** | 3.29*** | 11.3*** | 10.5*** | 9.65*** | 1.54 | 0.23 | 7.81* | 6.78*** | 2.10*** | 3.12*** |
| gdpcdif | 0.03 | 0.08 | 0.08 | -0.00 | 0.00 | 0.01 | 0.04 | 0.01 | -0.06 | -0.03 | -0.04 | -0.08*** |
| gdpg | 0.05 | 0.06 | 0.10* | -0.06 | -0.13 | -0.081 | 0.04 | 0.12** | 0.08 | 0.03 | 0.11 | 0.13 |
| popav | 1.06 | 0.26 | 0.00 | 32.9*** | 5.96*** | 5.65*** | -0.70 | -1.68** | 2.67 | -4.92 | 0.49 | 1.29 |
| bexr | -1.21*** | -0.24* | -0.28** | -0.25 | -0.33* | -0.24** | -0.59*** | -0.06*** | 0.06 | 0.31 | -0.62*** | -0.75*** |
| dist | - | -1.63*** | -1.67*** | - | -0.95*** | -0.96*** | - | -0.53*** | -0.25 | - | -0.61*** | -0.56*** |
| dummy | - | 1.64*** | 1.46*** | - | 0.57 | 0.62* | - | 0.71*** | 3.57** | - | 0.59 | 0.30 |
| Hausman test (recommended) | $P > \chi^2 = 0.00$ (Fixed) | | | $P > \chi^2 = 0.00$ (Fixed) | | | $P > \chi^2 = 0.26$ (Random) | | | $P > \chi^2 = 0.00$ (Fixed) | | |
| Pagan-Hall test | | 0.36 (1.00) | | | 25.4 (0.00) | | | 0.06 (1.00) | | | 0.21 (1.00) | |
| Wu-Hausman test | | 2.87 (0.04) | | | 1.56 (0.20) | | | -6.38 (1.00) | | | 5.64 (0.00) | |
| Durbin-Wu-Hausman test | | 9.21 (0.02) | | | 5.22 (0.15) | | | -31.8 (1.00) | | | 16.3 (0.00) | |
| Sargan test / Hansen J test | | 1.99 (0.36) | | | 4.36 (0.11) | | | 6.18 (0.04) | | | 6.71 (0.03) | |
| Observations | 84 | 84 | 83 | 84 | 84 | 83 | 84 | 84 | 83 | 84 | 84 | 83 |

N*otes: (i) \*, \*\*, \*\*\* means significance at 10 %, 5 % et 1 %; (ii) p-values are reported in brackets; (iii) The instruments used for the endogenous variables (outfdi, gdpg and bexr) are the first lags of these variables; (iv) Sargan statistic represents the over-identification test for all instruments and equals 0.00 if the equation is exactly identified. However, if the Pagan-Hall test documents the presence of heteroskedasticity, the Hansen J is recommended for identifying over-identification issues; (v) 2SLS$^r$ means 2SLS with robust errors if the heteroskedasticity is present; (vi) for endogeneity issues we resort to the Wu-Hausman and Durbin-Wu-Hausman tests; (vii) the dummy variable takes value1 for CEE-3 as partners, and 0 otherwise.*

The Pagan-Hall test signals heteroskedasticity concerns for all the panels. Therefore, we estimate the 2SLS model with robust standard errors. The Hansen J test shows in general that there are no over-identification concerns regarding the instruments, except for the Czech Republic.

After exploring the inward FDI – exports relationship, we are particularly interested in the influence of inward FDI on imports. If the inward FDI stock stimulates imports, this could be a sign that fairly large volumes of the parent company's intermediate products flow into the subsidiary's output, or that the foreign firms draw on established business relationships with foreign partner firms for intermediate products. Another explanation could be that the aim of the direct investments of the parent company is to acquire better access to the markets of the host country (Herrmann and Jochem, 2005). Table 8 presents the results.

**Table 8. Imports and inward FDI relationship**

| *imports* | Czech Republic | | | Hungary | | | Poland | | | Slovak Republic | | |
|---|---|---|---|---|---|---|---|---|---|---|---|---|
| | FE | RE | 2SLS$^r$ | FE | RE | 2SLS | FE | RE | 2SLS$^r$ | FE | RE | 2SLS$^r$ |
| *c* | -103** | -28.2** | -39.3*** | -543*** | -96.9*** | - | -12.1 | -20.2 | -0.55 | 51.8 | -43.3*** | -39.2 |
| **infdi** | 0.04 | 0.09* | 0.00 | 0.05 | 0.11*** | 0.35 | 0.04 | 0.04 | 0.30 | 0.16*** | 0.17*** | 0.14 |
| *gdpav* | -1.93 | -0.45 | -0.72 | -3.44** | -2.84 | 2.82** | 1.69 | 2.46*** | 2.00*** | -2.13** | 0.56 | 0.84 |
| *gdpdif* | 0.19 | 0.64*** | 0.66*** | -0.30 | 0.44*** | 0.15 | -0.10** | -0.10** | -0.11** | -1.13*** | -0.93*** | -0.98* |
| *gdpcav* | 5.41*** | 4.27*** | 5.30*** | 10.1*** | 8.20*** | 1.05 | 2.85*** | 1.78*** | 0.28 | 6.03*** | 2.63*** | 2.38** |
| *gdpcdif* | -0.01 | -0.12 | -0.15 | -0.01 | -0.00 | 0.15 | 0.03 | 0.00 | 0.04 | -0.04 | -0.00 | -0.02 |
| *gdpg* | 0.06 | 0.02 | 0.03 | -0.06 | -0.14 | 1.42 | 0.04 | 0.11* | 0.09* | 0.04 | 0.11 | 0.13** |
| *popav* | 5.48* | 0.48 | 0.75 | 30.6*** | 4.10** | -2.00 | -1.32 | -0.63 | -0.51 | -3.50 | 2.43*** | 2.16 |
| *bexr* | -1.39*** | -0.26* | -0.23* | -0.28 | -0.40** | -0.34 | -0.57** | -0.04 | -0.13* | 0.08 | -0.62*** | -0.63*** |
| *dist* | - | -1.49*** | -1.48*** | - | -0.93*** | -0.95*** | - | -0.56*** | -1.03** | - | -0.60*** | -0.59*** |
| *dummy* | - | 1.75*** | 1.77*** | - | 0.76 | 1.69 | - | 1.38*** | 1.43*** | - | 0.40 | 0.29 |
| Hausman test (recommended) | $P > \chi^2 = 0.00$ (Fixed) | | | $P > \chi^2 = 0.00$ (Fixed) | | | $P > \chi^2 = 0.40$ (Random) | | | $P > \chi^2 = 0.00$ (Fixed) | | |
| Pagan-Hall test | | | 0.65 (1.00) | | | 0.88 (0.99) | | | 0.46 (1.00) | | | 0.21 (1.00) |
| Wu-Hausman test | | | 1.19 (0.31) | | | 1.21 (0.31) | | | 1.45 (0.23) | | | 5.70 (0.00) |
| Durbin-Wu-Hausman test | | | 4.08 (0.25) | | | 4.09 (0.25) | | | 4.92 (0.17) | | | 16.4 (0.00) |
| Sargan test / Hansen J test | | | 1.28 (0.52) | | | 0.00 | | | 3.78 (0.15) | | | 5.98 (0.05) |
| Observations | 84 | 84 | 83 | 84 | 84 | 83 | 84 | 84 | 83 | 84 | 84 | 83 |

Notes: (i) *, **, *** means significance at 10 %, 5 % et 1 %; (ii) p-values are reported in brackets; (iii) The instruments used for the endogenous variables (infdi, gdpg and bexr) are the first lags of these variables; (iv) Sargan statistic represents the over-identification test for all instruments and equals 0.00 if the equation is exactly identified. However, if the Pagan-Hall test documents the presence of heteroskedasticity, the Hansen J is recommended for identifying over-identification issues; (v) 2SLS$^r$ means 2SLS with robust errors if the heteroskedasticity is present; (vi) for endogeneity issues we resort to the Wu-Hausman and Durbin-Wu-Hausman tests; (vii) the dummy variable takes value1 for CEE-3 as partners, and 0 otherwise.

In general, a positive relationship is documented, except for Poland, where none of the retained models validates the inward FDI – imports link. However, this complementarity effect documented *inter-alia* by Fillat-Castejón et al. (2008) is not supported by all the models and then, the results lack in robustness. There is a significant difference between the CEE-3 and EU-3 as partners, only in the case of the Czech Republic and Poland. In this case also, the average GDP per capita positively influences the imports, showing that more developed countries are more integrated.

To conclude, we can state that trade and FDI are complements rather that substitutes, in all the cases. We notice that investments in the partner countries enhance competitiveness on the foreign markets and conduct to a vertical integration and to a reallocation of labour-intensive activities from a capital-intensive country. However, the impact of the inward FDI on trade is less obvious. For all the CEE-4 countries, in all the cases, the trade – FDI nexus is more intense when considering the CEE countries as partners. Consequently, the bilateral trade between the CEE countries is influenced to a larger extent by the bilateral FDI, as compared to the bilateral trade between the CEE countries and the main EU commercial partners.

We admit however several limits of our estimations, generated especially by the small data sample and by possible manifestations of muticollinearity problems, given the characteristics of our control variables. Even if our estimations look robust, we only consider linear models. However, the crisis appearance may have caused structural breaks in the analysed relationships, but the lack of sufficient data for the CEE countries makes such analysis difficult for the moment (i.e. the consideration of time-dummy variables might represent a solution, but it leads to the weak instruments problem in the case of 2SLS estimation). In addition, we did not consider the role of business cycles correlation between the CEE countries. Finally, we did not examine the trade and investment's structure. A general picture of this nexus does not bring sufficient information about the role of FDI in promoting trade in the CEE countries. A sectorial analysis at firm level could lead to more diversified results, depending on the characteristics of the firms or industry in question.

## 5. Conclusions and policy implications

This paper aims to study the relationship between the bilateral trade and FDI in the case of four CEE countries, namely the Czech Republic, Hungary, Poland and the Slovak Republic, drawing a comparison between the role of bilateral investment between the CEE historical commercial partners and the main EU partners, namely Austria, Germany and Netherlands .

The trade – FDI nexus has been a topic of serious interest during the last decades, both for the trade and FDI theories, though it has generated different empirical results. Lately, these two different views emerged, accepting a complementarity effect, as well as a substitution effect between trade and FDI. However, at macro level, the complementarity effect prevails according to previous studies. Even if the subject is of great interest for the EU, little empirical evidence appears for the CEE countries. The few papers approaching the case of CEE countries investigate the role of FDI in the bilateral trade between these countries and the old EU members. None of these studies pays attention to the bilateral trade between the CEE countries, which knew great expansion in the period before the crisis and slowed down afterwards. Our paper is meant to fill in this gap and to analyse the relationship between the bilateral trade and FDI, from the perspective of each reporting country retained in the analysis.

We build a panel for each analysed country and we test four different cross-relationships which emerge between the outward/inward FDI and exports/imports. For robustness purposes and also for dealing with reverse causality issues, we compare a panel fixed effect and a random effect model, and we resort to a 2SLS estimation. We find a complementarity effect between trade and FDI, for all the investigated countries, which manifests especially in the case of the outward FDI. Moreover, the complementarity effect is stronger when we consider the CEE countries as partners. Therefore, vertical FDI outflows toward the historical trade partners encourage the commercial integration of the considered CEE countries.

These findings have several policy implications. First, because the inward FDI has no considerable influence on trade, the efforts of the authorities to attract FDI in order to support an increased commercial integration will be in vain. Second, we notice that the outward FDI favours both exports and imports, supporting thus the commercial integration of the CEE countries. These phenomena are stronger in the case of the CEE countries' bilateral investment. Consequently, in order to foster the commercial integration, the CEE countries' outward FDI shall be oriented toward their historical partners.

Two reasonable questions arise in this moment: (i) Toward which sectors shall the outward FDI be oriented? and (ii) How can it be encouraged? The response to the first question is complex. On the one hand, the outward FDI influences exports, owing to enhanced competitiveness on foreign markets. These findings characterise in particular the Czech and the Slovak Republic which exploit to a larger extent the local labour force of the CEE partner countries, and their natural resources, in order to increase the export levels. On the other hand, the outward FDI influences imports. This proves a backward vertical integration and a relocation of labour-intensive activities abroad. Regarding the theoretical considerations, our results are not surprising for the Czech Republic, but are questionable in the case of Poland, a country which is more labour-intensive than its CEE partners. In order to answer to the second question, we may consider the implementation of trade agreements. However, given the absence of trade barriers inside the EU and the cultural linkages of CEE-4, it is hard to consider new trade agreements as a viable solution. Nevertheless, increased transportation costs or euro adoption by the Czech Republic, Hungary and Poland, can favour the outward FDI.

All in all, the results indicate a complementarity effect between trade and FDI, which manifests in particular for the outward FDI and which is stronger between the CEE countries and their historical partners, as compared to the actual main EU partners. Nevertheless, more detailed information regarding the trade – FDI nexus can be obtained analysing the bilateral trade and FDI by industries, which can represent a subject for future developments of the present research.

# Appendixes

## Appendix A. FDI statistics

**Fig 2 Bilateral inward and outward FDI between each CEE-4 country and the CEE-3 (mil. US dollars)**

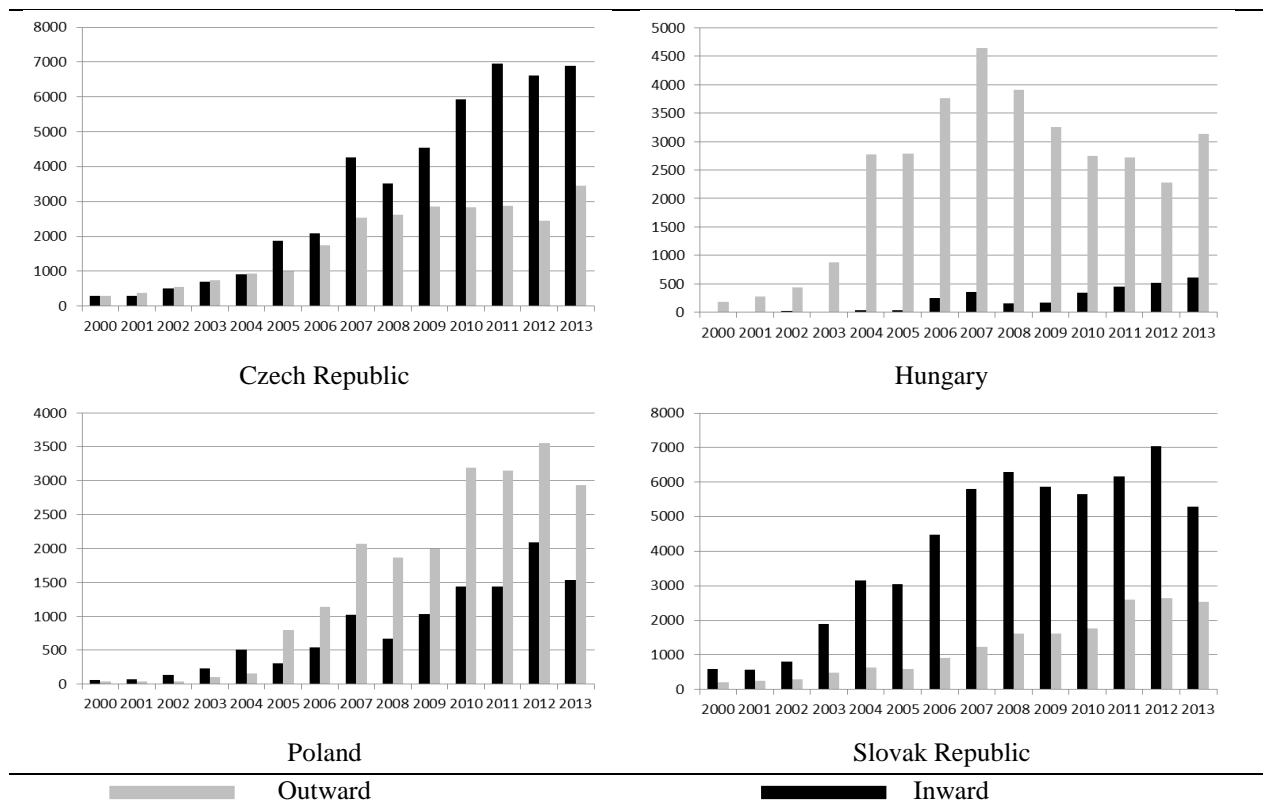

*Source: own calculations based on the OECD statistics*

## Appendix B. Explanatory variables

### Table 1. Explanatory variables description

| Variables | Description | Exp. sign | Employed in the literature by | Database |
|---|---|---|---|---|
| $outfdi_{ijt}$ | The volume of outward FDI from country $i$ to country $j$ at moment $t$. If FDI creates trade, then a positive influence is expected. The substitution is associated with a negative sign. | + / - | Pain and Wakelin (1998), Fontagné (1999) | OECD |
| $infdi_{ijt}$ | The volume of inward FDI from country $j$ to country $i$ at moment $t$. If FDI create trade, then positive influence is expected (complementarity relationship). The substitution relationship is associated with a negative sign. | + / - | Pain and Wakelin (1998), Fontagné (1999) | OECD |
| $gdpav_{ijt}$ | The average GDP volume of the countries $i$ and $j$ at moment $t$. It represents a *proxy* for the factors' income. | + | Africano and Magalhães (2005) | OECD |
| $gdpdif_{ijt}$ | The difference between the GDP volume of country $i$ and $j$ at moment $t$, in module. It represents a *proxy* for the difference of consumer preferences of the declaring country compared to its partner's. | + | Fontagné and Pajot (1999), Chiappini (2013) Sapienza (2009) | OECD |
| $gdpcpav_{ijt}$ | The average income per capita of the declaring country $i$ and its partner $j$ at moment $t$. It expresses the demand for variety. | + | Africano and Magalhães (2005) | OECD |
| $gdpcpdif_{ijt}$ | The differences between the GDP per capita in country $i$ and country $j$ at moment $t$. It represents the difference in factors endowments in the declaring country and in its partner country. It is measured in module. Depending on the structure of the relationship between trade and FDI, the expected sign can be either positive or negative. | +/- | Zarotiadis and Mylonidis (2005), Chiappini (2013) | OECD |
| $gdpg_{jt}$ | The GDP growth rate in the partner country $j$ at moment $t$. The economic growth of the partner country favours the intensification of trade flows with this country. Because the values of the indicators are negative in 2009, and because all the variables were transformed in logarithm, we employ the formula: $gdpg = 10\% +$ real growth rate. | + | Cheng and Wall (2005) | OECD |
| $popav_{ijt}$ | The average population of the two countries at moment $t$, associated with a big market favouring the trade. However, trade openness is much higher in small CEECs. Thus, either a positive or a negative sign is expected. | +/- | Fontagné and Pajot (1999), Kyrkilis and Pantelidis(2000) | OECD UNCTAD |
| $bexr_{ijt}$ | The bilateral exchange rate of the declaring country $i$ with its partner $j$ at moment $t$. An appreciation of the domestic currency favours imports and has a negative impact upon exports. | - / + | Carter and Yilmaz (1999), Kutan and Vukšić (2007) | Central banks statistics |
| $dist_{ij}$ | The geographical distance between the country $i$ and country $j$ (the distance between the capital cities of two countries). It represents a *proxy* for transportation costs. If the distance is high, the transportation costs negatively affect trade creation. | - | Cheng and Wall (2005), Africano and Magalhães (2005) | Via Michelin |
| $dummy$ | Takes value 1 if the partner country is a CEEC, while taking value 0 for ole EU members. | + | - | - |

*Notes: The GDP related variables are expressed in constant prices (US $).*